\documentclass[submission,copyright,creativecommons]{eptcs}
\providecommand{\event}{EXPRESS 2018} 

\usepackage{breakurl}             
\usepackage{underscore}           
\usepackage{microtype}
\usepackage{amssymb}
\usepackage{graphicx}
\usepackage{listings}
\usepackage{color}
\usepackage{rotating}
\usepackage{todonotes}
\usepackage{mathpartir}
\usepackage{url}
\usepackage{tikz}
\usepackage{amsmath}
\usepackage{stmaryrd}
\usepackage{amsthm}

\usetikzlibrary{calc,decorations.pathmorphing,shapes}
\newcounter{sarrow}
\newcommand\rlts[1]{%
\stepcounter{sarrow}%
\mathrel{\begin{tikzpicture}[baseline= {( $ (current bounding box.south) + (0,-0.5ex) $ )}]
\node[inner sep=.5ex] (\thesarrow) {$\scriptstyle #1$};
\path[draw,<-,decorate,
  decoration={zigzag,amplitude=1.2pt,segment length=1.5mm,pre=lineto,pre length=6pt}]
    (\thesarrow.south east) -- (\thesarrow.south west);
\end{tikzpicture}}%
}

\newtheorem{definition}{Definition}
\newtheorem{remark}{Remark}
\newtheorem{example}{Example}
\newtheorem{lemma}{Lemma}
\newtheorem{corollary}{Corollary}
\newtheorem{theorem}{Theorem}

\title{A Parametric Framework for Reversible $\pi$-Calculi\footnote{This work was partially supported by COST Action IC1405 ``Reversible computation -- extending horizons of computing", EPSRC EP/K034413/1, EP/K011715/1, EP/L00058X/1, EP/N027833/1 and 
EP/N028201/1.}}
\author{Doriana Medic
\institute{IMT School for Advanced Studies Lucca, Italy}
\email{doriana.medic@imtlucca.it}
\and
Claudio Antares Mezzina
\institute{IMT School for Advanced Studies Lucca, Italy}
\email{claudio.mezzina@imtlucca.it}
\and
Iain Phillips
\institute{ Imperial College London, UK}
\email{i.phillips@imperial.ac.uk}
\and
Nobuko Yoshida
\institute{Imperial College London, UK}
\email{n.yoshida@imperial.ac.uk}}
\def\titlerunning{A Parametric Framework for Reversible $\pi$-Calculi}
\def\authorrunning{D. Medic, C.A. Mezzina, I.C.C. Phillips \& N. Yoshida}
\def\copyrightholders{Medic, Mezzina, Phillips \& Yoshida}

\newcommand{\ltss}[2]{\xrightarrow[#1]{#2}}
   
\newcommand{\nil}{\mathbf{0}}
\newcommand{\para}{\;|\;}
\newcommand{\res}{\nu}
\newcommand{\mem}[1]{\lbrack{#1}\rbrack}
\newcommand{\g}{\Gamma}
\newcommand{\de}{\Delta}     
\newcommand{\out}[2]{\overline{{#1}}{#2}}
\newcommand{\outbo}[3]{\overline{{#1}}\langle\nu {#2}_{#3}\rangle}  
\newcommand{\lts}[1]{\xrightarrow{#1}}
\newcommand{\flts}[2][]{%
  \xrightarrow[#1]{#2}\mathrel{\mkern-14mu}\rightarrow}  
\newcommand{\trans}[4]{({#1},{#2},{#3}):{#4}} 
\newcommand{\sub}[2]{\{{#1}/{#2}\}}   

\newcommand{\ins}[1]{^{#1}}   
\newcommand{\outi}[2]{\overline{{#1}}^j{#2}}
\newcommand{\outin}[3]{\overline{{#1}}^{#3}{#2}}   

\newcommand{\pj}{\bold{P}}

\newcommand{\str}{\sqsubseteq}    
\newcommand{\stru}{\sqsubset}    
\newcommand{\obj}{\ll}   
\newcommand{\ob}{<} 
\newcommand{\cau}{\prec}    
\newcommand{\insta}{\rightsquigarrow}
\newcommand{\eq}{\sim}         

\newcommand{\rev}{^\bullet}

\newcommand{\rem}{_{\#i}}

\newcommand{\om}{\Omega}

\newcommand{\f}{\varphi}

\newcommand{\revprocs}{\mathcal{X}}

\newcommand{\act}{\mathcal{A}}
\newcommand{\lbl}{\mathcal{L}}

\newcommand{\upd}[2]{_{[{#1}/{#2}]\&i}}

\newcommand{\proc}{\triangleright}

\newcommand{\sdef}{::=}

\newcommand{\eventt}[3]{\langle #1,#2,#3\rangle }

\newcommand{\entry}[1]{\langle #1 \rangle}

\newcommand{\branch}{\uparrow}

\newcommand{\ids}{\mathcal{K}}
\newcommand{\names}{\mathcal{N}}
\newcommand{\vars}{\mathcal{V}}

\newcommand{\init}{\mathtt{init}}
\newcommand{\empt}{\mathtt{empty}}
\newcommand{\bool}{\mathtt{bool}}
\newcommand{\fresh}{\mathtt{fresh}}
\newcommand{\key}{\mathtt{key}}

\newcommand{\caus}[1]{\mathtt{Cause}(#1)}

\newcommand{\upda}[1]{\mathtt{Update}(#1)}

\newcommand{\hist}{\mathtt{H}} 
\begin{document}

\maketitle

\begin{abstract}
This paper presents a study of causality in a reversible, concurrent setting. 
There exist various notions of causality in $\pi$-calculus, which differ in the treatment of parallel extrusions of the same name. 
In this paper we present a uniform framework for reversible $\pi$-calculi that is \emph{parametric} with respect to a data structure that stores information about an extrusion of a name. 
Different data structures yield different approaches to the parallel extrusion problem.
We map three well-known causal semantics into our framework.
We show that the (parametric) reversibility induced by our framework is causally-consistent and prove a
causal correspondence between an appropriate instance of the framework and Boreale and Sangiorgi's causal semantics. 
 \end{abstract}
 \section{Introduction}
%
Starting from the 1970s~\cite{Bennett73} reversible computing has attracted interest in different fields, 
from thermodynamical physics~\cite{BacciDK11}, to systems biology~\cite{DanosK07a,PhillipsUY13}, system debugging~\cite{Zelkowitz,fase} and quantum computing~\cite{quantum}.
Of particular interest is its application to the study of programming abstractions for reliable systems: most fault-tolerant schemes exploiting system recovery techniques~\cite{depsys} rely on some form of \textit{undo}. 
Examples of how reversibility can be used to model transactions exist in CCS~\cite{DanosK05} and higher-order $\pi$-calculus~\cite{LaneseLMSS13}.

A reversible system is able to execute both in the forward (normal) direction and in the backward one. In a sequential setting, there is just one order of reversing a computation: one has just to undo the computation by starting from the last action. In a concurrent system  there is no clear notion of last action. A good approximation of what is the last action in a concurrent system 
is given by \textit{causally-consistent} reversibility, introduced by  Danos and Krivine for reversible CCS~\cite{rccs}. 
Causally-consistent reversibility relates causality and reversibility of a concurrent system in the following way: an action can be reversed, and hence considered as a last one, provided all its consequences have been reversed.

In CCS~\cite{CCS}, there exists just one notion of causality: so-called \textit{structural} causality, which is induced by the prefixing `.' operator and by synchronisations. As a consequence, there is only one way of reversing a CCS trace, and from an abstract point of view there exists only one reversible CCS.
Evidence for this has been given in~\cite{MedicM16}, where an equivalence is shown between the two methods for reversing CCS 
(namely RCCS~\cite{rccs} and CCSK~\cite{CCSK}).

When moving to more expressive calculi with name creation and value passing like the $\pi$-calculus, matters are more complex.
As in CCS, structural causality in the $\pi$-calculus is determined by the nesting of the prefixes; for example, in process $\out{b}{a}.\out{c}{e}$ the output on channel $c$ structurally depends on the output on~$b$. Extruding (or opening) a name generates an \textit{object} dependency; for example, in process
$\res{a} \;(\out{b}{a} \para a(z)) $ the input action on $a$ depends on the output on $b$. 
In the case of parallel extrusions of the same name, for example $\res{a} \;(\out{b}{a} \para \out{c}{a} \para a(z))$, 
there exist different interpretations of which extrusion will cause the action~$a(z)$. 
In what follows, we consider three approaches.

The classical and the most used approach to causality in the $\pi$-calculus is the one where the order of extrusions matters and
the first one of them is the cause of the action $a(z)$. Some of the causal semantics representing this idea are~\cite{DeganoP99,BorealeS98,BusiG95} and all of them are defined for standard (forward-only) $\pi$-calculus. 
In~\cite{DeganoP99} the authors claim that, after abstracting away from the technique used to record causal dependences, the final order between the actions in their semantics coincides with the ones introduced in~\cite{BorealeS98,BusiG95}.
Hence we group these semantics together as a single approach to causality.

Secondly, in~\cite{CrafaVY12}, action $a(z)$ in the example above depends on one of the extruders, but there is no need to keep track of which one exactly. This causal semantics is defined for the forward-only $\pi$-calculus.

Finally, the first compositional causal semantics for the reversible $\pi$-calculus is introduced in~\cite{CristescuKV13}. In the above example, parallel extrusions are concurrent and the action $a(z)$ will record dependence on one of them (exactly which one is decided by the context). This causal semantics enjoys certain correctness properties which are not satisfied by other semantics. 

Here we present a framework for reversible $\pi$-calculus that is parametric with respect to the data structure that stores information about an extrusion of a name. Different data structures will lead to different approaches to the parallel extrusion problem, including the three described above.
Our framework allows us to add reversibility to semantics where it was not previously defined. 
By varying the parameters, different orderings of the causally-consistent backward steps are allowed. 
Our intention is to develop a causal behavioural theory for the framework, in order to better understand different interpretations of reversibility in the $\pi$-calculus, and to use this understanding for causal analysis of concurrent programs.



A preliminary discussion of the framework appeared in~\cite{MedicM17}, where some initial ideas
were given. Moreover  in~\cite{MedicM17} it was argued that it was necessary to modify the semantics of~\cite{BorealeS98} in order to add information about silent actions. In this work we fully develop the idea behind the framework and  
leave the semantics of~\cite{BorealeS98} unchanged, apart from using a late semantics, rather than early as originally given.
%

\noindent\textbf{Contributions.} We present a framework for reversible $\pi$-calculus which is parametric in the bookkeeping data structure used to keep track of object dependency. As reversing technique, we will extend the one  introduced by CCSK~\cite{CCSK}, which is limited to calculi defined with GSOS~\cite{gsos} inference rules (e.g., CCS, CSP), to work with more expressive calculi featuring name passing and binders. This choice allows us to have a compositional semantics which does not rely on any congruence rule (in particular the splitting rule used by~\cite{CristescuKV13}).
Depending on the bookkeeping data structure used to instantiate the framework, we can obtain different causal semantics (i.e.,~\cite{CristescuKV13,BorealeS98,CrafaVY12}). We then show that our framework enjoys the standard properties for a reversible calculus, namely the loop lemma and causal consistency, regardless of the notion of causality which is used. We prove causal correspondence between the causal semantics introduced in~\cite{BorealeS98} and the
matching
instance of our framework. 

The rest of the paper is as follows:
 syntax and operational semantics of the framework are given in Section~\ref{sc:framework}. In Section~\ref{sc:mapping}, we show how by using different data structures we can encompass different causal semantics.
 The main results are given in Section~\ref{sc:results}, and Section~\ref{sc:conc} concludes the paper. Proofs are omitted for space reasons; they can be found in the extended version~\cite{MMPY18}.

\section{The Framework}\label{sc:framework}

We present the syntax and operational semantics of our parametric framework, after an informal introduction.
\subsection{Informal presentation}\label{ssc:informal}
In~\cite{CCSK} a general technique to reverse any CCS-like calculus is given. The key ideas are to use communication keys to identify events, and to make static all the operators of the calculus, since dynamic operators such as choice and  prefix are \emph{forgetful} operators. For example, if we take a CCS process
$a.P \para \overline{a}.Q$ a possible computation is: 

\vspace{-3mm}
$\qquad \qquad \qquad \qquad \qquad \qquad \qquad a.P \para \overline{a}.Q   \xrightarrow{\tau[i]} a[i].P \para \overline{a}[i].Q$
\vspace{1mm}

\noindent As one can see, prefixes are not destroyed but \textit{decorated} with a communication key. The obtained process acts like $P \para Q$, since decorated prefixes are just used for backward steps. We bring this idea to the $\pi$-calculus. For example 
by lifting this process into $\pi$-calculus we have something like
\vspace{-2mm}

\vspace{3mm}
$\qquad\qquad\qquad\qquad\quad\qquad  a(x).P \para \out{a}b.Q   \xrightarrow{\tau:i} a(x)[i].P\sub{b^i}{x} \para \overline{a}b[i].Q$
\vspace{2mm}

\noindent In the substitution $\sub{b^i}{x}$, name $b$ is decorated with the key $i$ to record that it was substituted for variable $x$ in the synchronisation identified by the communication key $i$. The key $i$ is also recorded in the memories. 

By choosing to adapt the ideas of~\cite{CCSK} to work with the $\pi$-calculus, we avoid using the splitting rule of $R\pi$~\cite{CristescuKV13}. In $R\pi$ each process is monitored by a memory, $m \proc P$, which is in charge of recording all past events of the process. In this way, the past of the process is not recorded directly in the process. One drawback of this approach is that one needs to resort to a splitting rule of the form 

\vspace{1mm}
$\qquad\qquad\qquad\qquad\qquad\quad\quad m \triangleright (P\para Q) \equiv \langle \uparrow\rangle \cdot m \triangleright P \para \langle \uparrow\rangle \cdot m \triangleright Q$
\vspace{1mm}

\noindent to let both $P$ and $Q$ execute. This rule is not associative and moreover, as shown in~\cite{rhojournal}, introduces some undesired non-determinism, since equivalent processes performing the same action may become non-equivalent processes.

The framework has to remember extrusions, and in particular  
who was the extruder of a certain name, and what is the \textit{contextual cause} for
an action.
For example in

\vspace{1mm}
$\qquad\qquad\qquad\res a \, (\out{b}{a} \para a(x).P) \xrightarrow{\outbo{b}{a}{}:i} \res a_{\{i\}}(\out{b}{a}[i] \para a(x).P)
\xrightarrow{a(x):j}\res a_{\{i\}}(\out{b}{a}[i] \para a(x)[j,i].P )$
\vspace{1mm}

\noindent we have that after the extrusion, the restriction $\res a$ does not disappear as in standard $\pi$-calculus, but remains where it was, becoming the memory $\res a_{\{i\}}$ (introduced in~\cite{CristescuKV13}). This memory records the fact that name $a$ was extruded because of transition $i$. Moreover, since it is no longer a restriction but just a decoration, the following transition using name $a$ can take place. Transition $j$ uses $i$ as its contextual cause, indicating that the input action can happen on $a$ because it was extruded by $i$, and this is recorded in the process $a(x)[j,i].P$.

\subsection{Syntax}\label{ssc:syntax}
We assume the existence of the following \textit{denumerable} infinite mutually disjoint sets: the set $\names$ of names, the set $\ids$ of keys, and the set $\vars$ of  variables. Moreover, we let $\ids_{*} = \ids \cup \{*\}$ where $*$ is a special key. We let $a,b,c$ range over $\names$; $x,y$ range over $\vars$ and $i,j,k$ range over $\ids$.

The syntax of the framework is depicted in Figure~\ref{fig:sys}. \textit{Processes},
 given by the $P, Q$ productions, are the standard processes of the $\pi$-calculus~\cite{sangiobook}:
 $\nil$ represents the idle process; $\out{b}{c}.P$
 is the 
 \textit{output}-prefixed process  indicating the act of sending name $c$ over channel $b$; 
 $b(x).P$ is the
 \textit{input}-prefixed process 
  indicating the act of receiving a value (which will be bound to the variable $x$) on channel $b$.
  Process $P \para Q$ represents the  \textit{parallel} composition of two processes, while $\res a (P)$
 represents the fact that name $a$ is \textit{restricted} in $P$.
\begin{figure}[t]
\vspace{-4mm}
\begin{align*}
X,\;Y\sdef \quad &\pj\para\; \outi{b}{a^{j_1}}\mem{i,K}.X\;\para\; b\ins{j}(x)\mem{i,K}.X\;\para\: X\para Y\;\para 
\res a_{\de} (X)\\
P,\;Q  \sdef \quad &\nil\para \out{b}{c}.P\para b(x).P \para \; P\para Q\; \para \res a (P)	 
\end{align*}
\vspace*{-7mm}
\caption{Syntax.}
\vspace{-2mm}
\label{fig:sys}
\end{figure}

 Reversibility is defined on the top of the $\pi$-calculus. Unlike in the standard $\pi$-calculus, executed actions are not discarded. Each of them, followed by the memory, becomes a part of the process that we shall call the \emph{history}.
 \textit{Reversible processes} are given by $X,Y$ productions. A reversible process $\pj$ is a standard $\pi$-calculus process $P$ where channels are decorated with instantiators. 
 As we shall see later on, instantiators are used to keep track of substitutions. 
 In a prefix of the form $\out{b}{a}$ or $b(x)$ we say that name $b$ is used in \textit{subject position}, while name $a$ and variable $x$ are in \textit{object position}. We shall use operators $sub(\cdot)$ and $obj(\cdot)$ to get respectively the subject and the object of a prefix.
   The prefix $\outi{b}{a^{j_1}}\mem{i,K}.X$ represents a \textit{past output} recording the fact that in the past the process $X$ performed an output identified by key $i$ and that its contextual cause set was $K\subseteq \ids_{*}$. Prefix 
   $b\ins{j}(x)\mem{i,K}.X$ represents a \textit{past input} recording the fact that 
   the input was identified by key $i$ and its contextual cause set was $K$.
   If it is not relevant whether the prefix in the process is an input or an output, we shall denote it with $\alpha$ ($\alpha=\outi{b}{a^{j_1}}$ or $\alpha=b^j (x)$).

    Following~\cite{CristescuKV13} the restriction operator 
 $\res a_{\de}$ is decorated with the memory $\de$
 which keeps track of the extruders of a name $a$. As we shall see later on, 
 we shall abstract away from the form of $\de$, as different data structures lead to different notions of causality.
 When $\empt(\de)=true$, the data structure is initialised and $\res a_{\de}$ will act as the usual restriction operator $\res a$ of the $\pi$-calculus.
The set of reversible processes is denoted with $\revprocs$.

 To simplify manipulation with reversible processes, we shall define history and general context. History context represents the reversible process $X$ made of executed prefixes. 
 For example, we can express the process $X=\outin{b}{a^*}{*}\mem{i,K}.\outin{c}{a^*}{*}\mem{i',K'}.\pj$  as $X=\hist[\pj]$ with $\hist[\bullet] = \outin{b}{a^*}{*}\mem{i,K}.\outin{c}{a^*}{*}\mem{i',K'}.\bullet$. 
 General context is defined on the top of the history context by adding parallel and restriction operators on it. For example, the process $Z\para Y\para X$ can be written as $C[X]$ if the only relevant element is $X$. 
 Formally:
 
 \begin{definition}[History and General context]
History contexts  $\hist$ and general contexts $C$ are reversible processes with a hole $\bullet$, defined by the following grammar:

\vspace{2mm}
$\qquad\qquad\qquad\qquad\hist::=\bullet\para\alpha\mem{i,K}.\bullet \qquad
C::=\hist[\bullet]\;\para\: X\para \bullet\;\para \res a_{\de} (\bullet)$
\end{definition}

\noindent\textbf{Free names and free variables.}
 Notions of free names and free variables in our framework are standard. 
 It suffices to note that constructs with binders are of the forms: 
 $\res{a}_\de (X)$ when $\empt(\de)$ holds, which binds the name $a$ with scope $X$; 
 and $ b(x).P $, which binds the variable $x$ with scope $P$. We denote with
 $\mathtt{fn}(P)$ and $\mathtt{fn}(X)$ the set of free names of $P$ and of $X$ respectively.

\begin{remark}
\normalfont{Annotation $b^*$ to a name $b$, used either in the subject or in the object position, indicates that name $b$ has no instantiators.}
\end{remark}
Since the framework will be parametric in the data structure $\de$, we specify it as an interface (in the style of a Java interface) by defining the operations that it has to offer. 

\begin{definition}\label{df:interface}
 $\de$ is a data structure with the following defined operations:
\begin{description}
\item[$(i)$] $\init: \de \rightarrow \de \, \text{ initialises the data structure}$
\item[$(ii)$] $\empt: \de \rightarrow \bool \, \text{ predicate telling whether $\de$ is empty}$
\item[$(iii)$] $+ : \de \times \ids \rightarrow \de \,  \text{ operation adding a key to $\de$} $
\item[$(iv)$]  $\#i : \de \times \ids \rightarrow \de \, \text{ operation removing a key from 
$\de$}$
\item[$(v)$] $\in : \de \times \ids \rightarrow \bool \, \text{ predicate telling whether a key belongs to $\de$}$
\end{description}
\end{definition}
We now define three instances of $\de$: sets, sets indexed with an element and sets indexed with a set.
As we shall see, these three instances will give rise to three different notions of object causality.

\paragraph*{\textbf{Set}.} $\g$ is a set containing keys (i.e.\ $\g \subseteq \ids$). The intuition of $\res a_\g$ is that \textbf{any} of the elements contained in $\g$ can be a contextual cause for $a$ (i.e., the reason why $a$ is known to the context). 
\begin{definition}[Operations on a set]
The operations on a set $\g$ are defined as: 
\begin{description} 
\item [$(i)$] $\init(\g)=\emptyset$
\item [$(ii)$] $\empt(\g)=true$, when $\g=\emptyset$
\item [$(iii)$] $+$ is the classical addition of elements to a set
\item [$(iv)$] $\#i$ is defined as the identity, that is 
$\de_{\rem}=\g_{\rem}=\g$.
\item [$(v)$] $i\in\g$ the key $i$ belongs to the set $\g$
\end{description}
\end{definition}
\paragraph*{\textbf{Indexed set}.}
$\g_w$ is an indexed set containing keys and $w$ is the key of the action which extruded a name~$a$.
In this case the contextual cause for name $a$ can be \textbf{just} $w$.
If there is no cause, then we shall set $w=*$. 
\begin{definition}[Operations on an indexed set]
The operations on an indexed set $\g_w$ are defined as:
\begin{description} 
\item [$(i)$] $\init(\g_w)=\emptyset_{*}$
\item [$(ii)$] $\empt(\g_w)=true$, when $\g=\emptyset \;\wedge\; w=*$
\item [$(iii)$] operation $+$ is defined as:
$
\g_w + i=
\begin{cases}
(\g\cup \{i\})_{i}, \text{ when }  w=*\\
(\g\cup \{i\})_{w}, \text{ when } w\neq *
\end{cases}
$
\item[$(iv)$] operation $\#i$ is defined inductively as:
\vspace{-2mm}
\begin{align*}
& (X\para Y)\rem= X\rem\para Y\rem  &&(\hist[X])\rem=\hist[X\rem] \qquad\qquad (\pj)\rem=\pj\\
& (\res a_{\g_{i}} X)\rem= \res a_{\g_{*}} X\rem &&
 (\res a_{\g_{w}} X)\rem= \res a_{\g_{w}} X\rem
\end{align*}
\vspace{-7.5mm}

\item [$(v)$] $i\in\g_w$ the key $i$ belongs to the set $\g$, regardless of $w$ (e.g. $i\in \{i\}_{*}$) 
\end{description}
\end{definition}
\paragraph*{\textbf{Set indexed with a set}.}
$\g_\om$ is a set containing keys indexed with a set $\om\in \ids_{*}$. 
Extruders of name $a$ which are not part of the communication, will be saved in the set $\om$. 
In this case the contextual cause for name $a$ is a set $\om$.
If there is no cause, then we shall set $\om=\{*\}$. 
\begin{definition}[Operations on a set indexed with a set]
The operations on a set indexed with a set $\g_{\om}$ are defined as:
\begin{description} 
\item [$(i)$] $\init(\g_{\om})=\emptyset_{\{*\}}$
\item [$(ii)$] $\empt(\g_{\om})=true$, when $\g=\emptyset \;\wedge\; \om=\{*\}$
\item [$(iii)$] operation $+$ is defined as:
$
(\g_{\om}) + i=(\g\cup \{i\})_{(\om\cup \{i\})}
$
\item [$(iv)$] operation $\rem$ is defined inductively as:

\vspace{1mm}
$\qquad (X\para Y)\rem= X\rem\para Y\rem \qquad(\res a_{\g_{\om}} X)\rem= \res a_{\g_{\om\setminus\{i\}}} X\rem\qquad(\hist(X))\rem=\hist[X\rem] \qquad (\pj)\rem=\pj$
\item [$(v)$] $i\in\g_{\om}$ the key $i$ belongs to the set $\g$, regardless $\om$ (e.g. $i\in \{i\}_{\{*\}}$) 
\end{description}
\end{definition}

\subsection{Operational Semantics}
The grammar of the labels generated by the framework is:
$$\mu::=\trans{i}{K}{j}{\pi}
\qquad \qquad\pi::=\out{b}{c}\para b(x)\para \outbo{b}{c}{\de}\para \tau$$
where $i$ is the key, and $K\subseteq \ids_{*}$, $j\in \ids_{*}$ are the set of contextual causes and an instantiator of $i$, respectively. If there is no action which caused and/or instantiated $i$, we denote this with $K=\{*\}$, $j=*$, respectively. 
 The set 
$\lbl$ of all possible labels generated by the framework is defined as $\lbl = \ids \times \ids_{*} \times \ids_{*} \times \act $, where $\act$ is a set of actions ranged over by $\pi$. We extend $sub(\cdot)$ and $obj(\cdot)$ to apply also to labels.

The operational semantics of the reversible framework is given in terms of a labelled transition system (LTS) $(\revprocs,\lbl,\lts{})$, where $\revprocs$ is the set of reversible processes; $\lts{}= \flts{} \cup  \rlts{\quad}$ where  $\flts{}$ is the least transition relation induced by the rules in 
Figures~\ref{fig:rulecommon} and~\ref{fig:ruleparam}; and $\rlts{\quad}$ is the least transition relation induced by the rules in Figure~\ref{fig:ruleback}.


\begin{definition}[Process keys]	
The set of \emph{communication keys} of a process $X$, written $\key(X)$, is inductively defined as follows:
\begin{align*}
&\key(X\para Y) = \key(X) \cup \key(Y)&& \key(\alpha\mem{i,K}.X) = \{i\} \cup \key(X) \\
& \key(\res a_{\de} (X)) = \key(X) && \key(\pj) =  \emptyset &
\end{align*}
\end{definition}
\begin{definition}
A key $i$ is \emph{fresh} in a process $X$, written $\fresh(i,X)$ if $i \not \in \key(X)$.
\end{definition}

The forward rules of a framework are divided into two 
\textit{groups}, depending on whether they are parametric with respect to $\de$ or they are common to all the instances of the framework.

\begin{figure}[t]
\vspace{-3mm}
{\footnotesize
\begin{mathpar}
\inferrule*[left=(\textsc{Out1})]
{}
{ \outi{b}{a^{j_1}}.\pj\flts{(i,K,j):\out{b}{a}}\outi{b}{a^{j_1}}\mem{i,K} .\pj }
\and
\inferrule*[left=(\textsc{Out2})]
{X\flts{\trans{i}{K}{j}{\out{b}{a}}} X'\quad \fresh(i,\hist[X])}
{\hist[X]\flts{\trans{i}{K}{j}{\out{b}{a}}}\hist[X'] }
\and
\inferrule*[left=(\textsc{In1})]
{}
{b\ins{j}(x).\pj\flts{(i,K,j):b(x)}b\ins{j}(x)\mem{i,K} .\pj}
\and
\inferrule*[left=(\textsc{In2})]
{X\flts{\trans{i}{K}{j}{b(x)}} X'\quad \fresh(i,\hist[X])}
{ \hist[X]\flts{\trans{i}{K}{j}{b(x)}} \hist[X'] }\\
\and
\inferrule*[left=(\textsc{Par})]
{X\flts{\trans{i}{K}{j}{\pi}}  X'\quad i\notin Y}
{ X\para Y\flts{\trans{i}{K}{j}{\pi}} X'\para Y }
\and
\inferrule*[left=(\textsc{Res})]
{X\flts{\trans{i}{K}{j}{\pi}} X'\quad a\notin\pi}
{  \res a_{\de} (X)\flts{\trans{i}{K}{j}{\pi}}\res a_{\de} (X') }
\and
\inferrule*[left=(\textsc{Com})]
{X\flts{\trans{i}{K}{j}{\out{b}{a}}} X'\quad Y\flts{\trans{i}{K'}{j'}{b(x)}}Y' \quad K=_{*}j'\;\wedge\;K'=_{*}j}
{ X\para Y\flts{\trans{i}{*}{*}{\tau}}X'\para Y'\sub{a^i}{x} }

\end{mathpar}
}
\vspace{-4mm}
\caption{Rules that are common to all instances of the framework.}
\label{fig:rulecommon}
\vspace{-1mm}
\end{figure}
Common rules are given in Figure~\ref{fig:rulecommon}.
Rules $\textsc{Out1}$ and $\textsc{In1}$ generate a fresh new key $i$ which is bound to the action. 
 Rules $\textsc{Out2}$ and $\textsc{In2}$ inductively allow a prefixed process $\hist [X]$ to execute if X can execute. 
 Condition $i\notin Y$ in rule $\textsc{Par}$ ensures that action keys are unique. Rule $\textsc{Res}$ is defined in the usual way. Two processes can synchronise through the rule $\textsc{Com}$ if the additional condition is satisfied ($K=_{*}j$ means $*\in K$ or $j=*$ or $K=j$). After the communication, necessary substitution is applied to the rest of the input process. In the process $Y'\sub{a\ins{i}}{x}$ every occurrence  of variable $x\in \mathtt{fn}(Y')$ is substituted with the name $a^{i}$, that is, the name $a$ decorated with the key $i$ of the action which was executed.  In the further actions of a process $Y'\sub{a\ins{i}}{x}$, the key $i$ will be called the \emph{instantiator}. The instantiators are used just to keep track of the substitution, not to define a name. For example, the two processes $\outi{b}{a^{*}}.\pj$ and $b^{j'}(x).\pj'$ can communicate, even if the instantiators of the name $b$ are not the same.
Let us note that we use a \textit{late} semantics, since substitution happens in the rule $\textsc{Com}$.
In order to understand how the basic rules work let us consider the following example.
\begin{example}\label{ex:rules}
Let $X=\outin{b}{a^{*}}{*}.\nil\para b^{*}(x).\out{x}{c^{*}}$. There are two possibilities for the process $X$:
\begin{itemize}
\item process $X$ can preform an output and an input action on the channel $b$ while synchronising with environment:
$$\outin{b}{a^{*}}{*}.\nil\para b^{*}(x).\out{x}{c^{*}}\flts{(i,*,*):\out{b}{a}}\outin{b}{a^{*}}{*}\mem{i,*}.\nil\para b^{*}(x).\out{x}{c^{*}}
\flts{(i',*,*):b(x)}\outin{b}{a^{*}}{*}\mem{i,*}.\nil\para b^{*}(x)\mem{i',*}.\out{x}{c^{*}}=Y_1$$
As we can notice, the output action $\out{b}{a}$ is identified by key $i$, while the input action is identified by key $i'$.
\item The synchronisation can happen inside of the process $X$:
$$\outin{b}{a^{*}}{*}.\nil\para b^{*}(x).\out{x}{c^{*}}\flts{(i,*,*):\tau}\outin{b}{a^{*}}{*}\mem{i,*}.\nil\para b^{*}(x)\mem{i,*}.\outin{a}{c^{*}}{i}=Y_2$$
We can notice that $\tau$ action is identified with key $i$ and during the synchronisation variable $x$ is substituted with a received name $a$ decorated with the key $i$ of the executed action. In this way we keep track of the substitution of a name.
\end{itemize}
\end{example}

We now define the operation $X\upd{K'}{K}$, which updates the contextual cause $K$ of an action identified by $i$ with the new cause $K'$. Contextual cause update will be used in the the parametric rules of Figure~\ref{fig:ruleparam} ($\textsc{Open}$ and $\textsc{Cause Ref}$). Formally:
\begin{definition} [Contextual Cause Update]
The \emph{contextual cause update} of a process, written $X\upd{K'}{K}$ is defined as follows:
{\small
\begin{align*}
& (X\para Y)\upd{K'}{K}= X\upd{K'}{K} \para Y\upd{K'}{K}&& \hist [\alpha[i,K].X]\upd{K'}{K}=\hist[\alpha [i,K']. X] &\\
& (\res a_\de (X))\upd{K'}{K}= \res a_\de (X)\upd{K'}{K}&& \hist [\alpha [j,K].X]\upd{K'}{K}=\hist[\alpha [j,K]. X] &
\end{align*}
}
\end{definition}

\begin{figure}[t]
{\footnotesize
\begin{mathpar}
\inferrule*[left=(\textsc{Cause Ref})]
{X\flts{\trans{i}{K}{j}{\pi}} X'\quad a\in sub(\pi)\qquad \empt(\de)\neq true \quad  \caus{\de,K,K'} }
{  \res a_{\de} (X)\flts{\trans{i}{K'}{j}{\pi}}\res a_{\de} (X'\upd{K'}{K}) }
\and
\inferrule*[left=(\textsc{Open})]
{X\flts{\trans{i}{K}{j}{\pi}} X'\quad \pi=\out{b}{a} \vee \pi=\outbo{b}{a}{\de'}\quad\upda{\de,K,K'} }
{  \res a_{\de} (X)\flts{\trans{i}{K'}{j}{\outbo{b}{a}{\de}}}\res a_{\de+i} (X'\upd{K'}{K}) } 
\and
\inferrule*[left=(\textsc{Close})]
{X\flts{\trans{i}{K}{j}{\outbo{b}{a}{\de}}} X'\quad Y\flts{\trans{i}{K'}{j'}{b(x)}}Y'\quad K=_{*}j'\;\wedge\;K'=_{*}j}
{ X\para Y\flts{\trans{i}{*}{*}{\tau}}\res a_{\de}(X'\rem\para Y'\sub{a^i}{x})}
\end{mathpar}
}
\vspace{-4mm}
\caption{Parametric rules}
\label{fig:ruleparam}
\vspace{-1mm}
\end{figure}

Parametric rules are given in Figure~\ref{fig:ruleparam}. Depending on the underlying causal semantics the way a contextual cause is chosen differs. This is why we need to define two predicates:
 $\caus{\cdot}$ and $\upda{\cdot}$. When instantiating $\de$ with a specific data structure, different implementations of such predicates are needed. We shall define them precisely when discussing various causal semantics in the Section~\ref{sc:mapping}.
Every time a label produced by a process $X$ passes the restriction $\res a_{\de}$ it needs to check if it is necessary to modify the contextual cause. 
Depending on whether name $a$ is in the subject or in the object position in the label of an action, rules $\textsc{Cause Ref}$ or $\textsc{Open}$ can be used, respectively.
Rule  $\textsc{Cause Ref}$ is used when the subject of a label is an already extruded name and
a predicate $\caus{\de,K,K'}$ tells whether contextual cause $K$ has to be substituted with $K'$. 
Rule $\textsc{Open}$ deals with the scope extrusion of a restricted name. If the restricted name $a$ is used as object of a label with key $i$ we have to record that $i$ is one of the potential extruders of $a$. Naturally, if $\empt(\de)=true$ then the first extruder initialises the data structure. Also in this case it might happen that we have to update the contextual cause of the label $i$. This is why predicate $\upda{\de,K,K'}$ is used. Two processes can synchronise through the rule $\textsc{Close}$ satisfying the additional condition. In some semantics, silent actions do not bring the causal information on what is a reason to introduce the operator $\rem$, where every time when an extruded name is closed over the context, the key of the closing action is deleted from indexes of $\de_h$ in all restrictions $\res a_{\de_h}\in X'$.

%
\begin{figure}[t]
\vspace{-3mm}
{\footnotesize
\begin{mathpar}
\inferrule*[left=(\textsc{Out1$\rev$})]
{}
{\outi{b}{a}\mem{i,K} .\pj\rlts{(i,K,j):\out{b}{a}} \outi{b}{a}.\pj}
\and
\inferrule*[left=(\textsc{Out2$\rev$})]
{X'\rlts{\trans{i}{K}{j}{\out{b}{a}}} X\quad \fresh(i,\hist [X])}
{ \hist[X']\rlts{\trans{i}{K}{j}{\out{b}{a}}}\hist [X] }
\and
\inferrule*[left=(\textsc{In1$\rev$})]
{}
{b\ins{j}(x)\mem{i,K} .\pj\rlts{\trans{i}{K}{j}{b(x )}} b\ins{j}(x).\pj}
\and
\inferrule*[left=(\textsc{In2$\rev$})]
{X'\rlts{\trans{i}{K}{j}{b(x)}} X\quad \fresh(i,\hist [X])}
{ \hist [X']\rlts{\trans{i}{K}{j}{b(x)}} \hist [X] }\\
\and
\inferrule*[left=(\textsc{Par$\rev$})]
{X'\rlts{\trans{i}{K}{j}{\pi}}  X\quad i\notin Y}
{ X'\para Y\rlts{\trans{i}{K}{j}{\pi}} X\para Y }
\and
\inferrule*[left=(\textsc{Com$\rev$})]
{X'\rlts{\trans{i}{K}{j}{\out{b}{a}}} X\quad Y'\rlts{\trans{i}{K'}{j'}{b(x)}}Y \quad K=_{*}j'\;\wedge\;K'=_{*}j }
{ X'\para Y'\rlts{\trans{i}{*}{*}{\tau}}X\para Y \sub{x}{a^{i}}}\\
\and
\inferrule*[left=(\textsc{Res$\rev$})]
{X'\rlts{\trans{i}{K}{j}{\pi}} X\quad a\notin\pi}
{  \res a_{\de} (X')\rlts{\trans{i}{K}{j}{\pi}}\res a_{\de} (X) }
\and
\inferrule*[left=(\textsc{Open$\rev$})]
{X'\rlts{\trans{i}{K}{j}{\pi}} X\quad \pi=\out{b}{a} \vee \pi=\outbo{b}{a}{\de'}\quad\upda{\de,K,K'} }
{  \res a_{\de+i} (X')\rlts{\trans{i}{K'}{j}{\outbo{b}{a}{\de}}}\res a_{\de} (X) } \\
\and
\inferrule*[left=(\textsc{Cause Ref$\rev$})]
{X'\rlts{\trans{i}{K}{j}{\pi}} X\quad a\in sub(\pi) \qquad \empt(\de)\neq true \quad  \caus{\de,K,K'} }
{  \res a_{\de} (X')\rlts{\trans{i}{K'}{j}{\pi}}\res a_{\de} (X) }
\and
\and
\inferrule*[left=(\textsc{Close$\rev$})]
{X'\rlts{\trans{i}{K}{j}{\outbo{b}{a}{\de}}} X\quad Y'\rlts{\trans{i}{K'}{j'}{b(x)}}Y\quad K=_{*}j'\;\wedge\;K'=_{*}j}
{ \res a_{\de}(X'\para Y')\rlts{\trans{i}{*}{*}{\tau}}X\para Y\sub{x}{a^{i}}}
\end{mathpar}
}
\vspace{-3mm}
\caption{Backward rules.}
\label{fig:ruleback}
\vspace{-1mm}
\end{figure}
Backward rules are symmetric to the forward ones; they are presented in Figure~\ref{fig:ruleback}. The predicates are not necessary for the backward transitions as they are invariant in the history of processes but we keep them to simplify the proofs. 
In order to better understand the backward rules, we shall consider the following example.
\begin{example} Let us consider the following processes from Example~\ref{ex:rules}:
\begin{itemize}
\item $Y_1=\outin{b}{a^{*}}{*}\mem{i,*}.\nil\para b^{*}(x)\mem{i',*}.\out{x}{c^{*}}$; Process $Y_1$ can perform backward actions on the channel $b$ (an input action identified with key $i'$ and an output action identified with key $i$) in any order. For example, let us reverse first the input and then the output action:
$$Y_1=\outin{b}{a^{*}}{*}\mem{i,*}.\nil\para b^{*}(x)\mem{i',*}.\out{x}{c^{*}}\rlts{\trans{i'}{*}{*}{b(x)}}
\outin{b}{a^{*}}{*}\mem{i,*}.\nil\para b^{*}(x).\out{x}{c^{*}}\rlts{\trans{i}{*}{*}{\out{b}{a}}}
\outin{b}{a^{*}}{*}.\nil\para b^{*}(x).\out{x}{c^{*}}=X$$
We notice that all the necessary elements to reverse the action $b(x)$ are saved in the history part of the process $Y_1$. 
\item $Y_2=\outin{b}{a^{*}}{*}\mem{i,*}.\nil\para b^{*}(x)\mem{i,*}.\outin{a}{c^{*}}{i}$; Process $Y_2$ can reverse the communication which happened on the channel $b$, between its subprocesses. Due to the side condition of the rule $\textsc{Par$\rev$}$, it is impossible to reverse an input or an output action separately:
$$\outin{b}{a^{*}}{*}\mem{i,*}.\nil\para b^{*}(x)\mem{i,*}.\outin{a}{c^{*}}{i}\not \rlts{\trans{i}{*}{*}{\out{b}{a}}}\outin{b}{a^{*}}{*}.\nil\para b^{*}(x)\mem{i,*}.\outin{a}{c^{*}}{i}$$
The backward action above cannot be executed as the key $i$ belongs to the process in parallel ($i\in\key(b^{*}(x)\mem{i,*}.\outin{a}{c^{*}}{i})$). The only possible backward step is:
$$\outin{b}{a^{*}}{*}\mem{i,*}.\nil\para b^{*}(x)\mem{i,*}.\outin{a}{c^{*}}{i} \rlts{\trans{i}{*}{*}{\tau}}
\outin{b}{a^{*}}{*}.\nil\para b^{*}(x).\out{x}{c^{*}}=X$$
\end{itemize}
\end{example}

\begin{remark} \normalfont{The choice operator $(+)$, can be easily added to the framework by following the approach of~\cite{CCSK} and by making the operator static.} 
\end{remark}

\section{Mapping causal semantics of $\pi$ into the framework}\label{sc:mapping}

We now review three notions of causal semantics for $\pi$-calculus and show how to map them into our framework by giving 
the definitions for the side conditions in the rules in Figure~\ref{fig:ruleparam}.

\paragraph*{\textbf{$R\pi$-calculus}.}
Cristescu et al~\cite{CristescuKV13} introduce a compositional semantics for the reversible $\pi$-calculus.  Information about the past actions is kept in a memory added to every process. A term of the form $m\proc P$ represents a reversible process, where memory $m$ is a stack of events and $P$ is the process itself. A memory contains two types of events, one which keeps track of the past action, $\eventt{i}{k}{\pi}$, where elements of a triple are the key, the contextual cause and the executed action, respectively; and one which keeps track of the position of the process in the parallel composition, $\entry{\branch}$. Before executing in parallel, a process splits by duplicating its memory and adding event $\entry{\branch}$ on the top of each copy. This is achievable with specially defined structural congruence rules.
The use in~\cite{CristescuKV13} of indexed restriction $\res a_{\g}$ was the inspiration for our parametric indexed restriction $\res a_{\de}$. A key point of the semantics of~\cite{CristescuKV13} is that it enjoys certain correctness properties: one of which is that two visible transitions are causally related iff for all contexts the corresponding silent transitions are.
Since an action can be caused only through the \textit{subject} of a label we have that contextual cause $K$ will be a singleton. We shall consider one relation between the prefixes into the history. In this way, while changing the cause with the rule $\textsc{Cause ref}$, the condition $\caus{\cdot}$ needs to keep track of the instantiation of the cause.

\begin{definition}[Instantiation relation]
Two keys $i_1$ and $i_2$ such that $i_1 , i_2\in\key(X)$ and $X=C[b^*(x)[i_1,K_1].Y]$ with $Y=C'[\alpha^{j_2}[i_2,K_2].Z]$ 
, are in instantiation relation, $i_1\insta_{X}i_2 $, if $j_2=i_1$.
If $i_1\insta_{X}i_2 $ holds, we will write $K_1\insta_{X}K_2 $.
\end{definition}

To obtain $R\pi$ causality in our framework, we need to instantiate the rules of
Figure~\ref{fig:ruleparam} with the following predicates.
\begin{definition} [$R\pi$ causality] If data structure $\de$ is instantiated with a set
$\g$, the predicates from Figure~\ref{fig:ruleparam} are defined as:
\begin{description}
\item [1.] $\caus{\de,K,K'}=\caus{\g,K,K'}$ stands for $K=K'$ or $\exists K'\in\g \; K\insta_{X}K'$;
\item [2.] $\upda{\de,K,K'}=\upda{\g,K,K'}$ stands for $K'=K$. 
\end{description}
\end{definition}
The predicates defined above coincide with the conditions of the semantics introduced in~\cite{CristescuKV13}.
In the following example we shall give the intuition of the $R\pi$ causality using our framework.
\begin{example}\label{ex:ex1}
Let us consider the process $X=\res a_\emptyset (\outin{b}{a^*}{*}\para \outin{c}{a^*}{*}\para a^{*}(x))$. By applying rule $\textsc{Open}$ twice and executing concurrently two extrusions on names $b$ and $c$, we obtain a process:
$$\res a_{\{i,h\}} (\outin{b}{a^*}{*}\mem{i,*}\para \outin{c}{a^*}{*}\mem{h,*}\para a^{*}(x))$$
The rule $\textsc{Cause Ref}$ is used for the execution of the third action. By definition of the predicate $\caus{\cdot}$,
the action $a(x)$ can choose its cause from a set $\{i,h\}$. By choosing $h$ for example, and executing the input action, we get the process:
$$\res a_{\{i,h\}} (\outin{b}{a}{*}\mem{i,*}\para \outin{c}{a}{*}\mem{h,*}\para a^{*}(x)\mem{l,h})$$
In the memory $\mem{l,h}$ we can see that the action identified with key $l$ needs to be reversed before the action with key $h$. Process $\outin{b}{a}{*}\mem{i,*}$ can execute a backward step at any time with the rule $\textsc{Open}\rev$.
\end{example}

\paragraph*{\textbf{Boreale-Sangiorgi and Degano-Priami causal semantics}.}

A compositional causal semantics for standard (i.e., forward only) $\pi$-calculus was introduced by Boreale and Sangiorgi~\cite{BorealeS98}. Later on, Degano and Priami in~\cite{DeganoP99} introduced a causal semantics for $\pi$ based on localities.
 While using different approaches to keep track of the dependences in $\pi$-calculus, these two approaches impose the same order of the forward actions (as claimed in~\cite{DeganoP99}). Hence, from the reversible point of view we can take it that the causality notions of these two semantics coincide.
 In what follows we shall concentrate on the Boreale-Sangiorgi causal semantics. 
 To show the correspondence between the mentioned semantics and our framework, we shall consider it in a late (rather than early, as originally given) version. The precise definition is given in~\cite{MMPY18}.

 The authors distinguish between two types of causality: \textit{subject} and the \textit{object}.
To capture the first one, they introduce a causal term $\mathtt{K}::A$, where $\mathtt{K}$ is a set of causes recording that every action performed by $A$ depends on $\mathtt{K}$. 
 The object causality is defined on the run (trace) of a process in such a way that once a bound name been extruded, it causes all the subsequent actions using that name in any position of the label. 
 Since an action can be caused through the \textit{subject} and \textit{object} position of a label, the contextual cause is a set $K\subseteq \ids_{*}$. For example, let us consider a process $\res a (\res b (\out{c}{b}\para \out{d}{a}\para \out{b}{a}))$ with a trace $\lts{\outbo{c}{b}{}}\lts{\outbo{d}{a}{}}\lts{\out{b}{a}}$. The action $\out{b}{a}$ depends on the first action because with it name $b$ was extruded and on the second action because with it name $a$ was extruded.
It is important to remark that a silent action does not exhibit causes.

To capture Boreale-Sangiorgi late semantics we need to 
give definitions for the predicates in Figure~\ref{fig:ruleparam}.

\begin{definition} [Boreale-Sangiorgi causal semantics] If an indexed set $\g_w$ is chosen as a data structure for a memory $\de$, the predicates from Figure~\ref{fig:ruleparam} are defined as:
\begin{description}
\item [1.] $\caus{\de,K,K'}=\caus{\g_{w},K,K'}$ stands for $K'=K\cup \{w\}$
\item [2.] $\upda{\de,K,K'}=\upda{\g_{w},K,K'}$ stands for $K'=K\cup \{w\}$
\end{description}
\end{definition}
Let us comment on the above definition. After the first extrusion of a name, the cause is fixed and there 
is no possibility of choosing another cause from the set $\g$. To capture this behaviour we
use the key of the first extruder, say $w$, as the index of the set $\g$.
The following example explains how our framework captures Boreale-Sangiorgi causality. We shall use the same process as in Example~\ref{ex:ex1}.
\begin{example}\label{ex:boreale}
Consider the process $X=\res a_{\emptyset_*} (\outin{b}{a^*}{*}\para \outin{c}{a^*}{*}\para a^{*}(x))$. By applying rule $\textsc{Open}$ and executing the first extrusion on name $b$, we obtain the process:
$$\res a_{\{i\}_{i}} (\outin{b}{a^*}{*}\mem{i,*}\para \outin{c}{a^*}{*}\para a^{*}(x))$$
In the memory ${\{i\}_{i}}$ the index $i$ indicates that name $a$ was extruded with the action $i$.
On the process $ \outin{c}{a^*}{*}$, rule $\textsc{Open}$ can be applied. By definition of the predicate $\upda{\cdot}$, the output action is forced to add $w=i$ in its cause set. Similar for the process $a^{*}(x)$, by applying the rule $\textsc{Cause Ref}$ and definition of the predicate $\caus{\cdot}$. After two executions, we obtain the process:
$$\res a_{\{i,h\}_{i}} (\outin{b}{a^*}{*}\mem{i,*}\para \outin{c}{a^*}{*}\mem{h,\{i,*\}}\para a^{*}(x)\mem{l,\{i,*\}})$$

In the memories $\mem{h,\{i,*\}}$ and $\mem{l,\{i,*\}}$ we see that both executed actions are caused by action $i$ and this is why it needs to be reversed last.
The second and the third action can be reversed in any order.
\end{example}

\paragraph*{\textbf{Crafa, Varacca and Yoshida causal semantics}.}
The authors introduced a compositional event structure semantics for the forward $\pi$-calculus~\cite{CrafaVY12}. They represent a process as a pair $(E,\mathtt{X})$, where $E$ is a prime event structure and $\mathtt{X}$ is a set of bound names. Disjunctive objective causality is represented in such a way that an action with extruded name in the subject position can happen if at least one extrusion of that name has been executed before. In the case of parallel extrusions of the same name, an action can be caused by any of them, but it is not necessary to remember which one.

Consequently, events do not have a unique causal history. As discussed in~\cite{rigidfamily} this type of disjunctive causality cannot be expressed when we consider processes with a contexts.
To adapt this notion of causality to reversible settings we need to keep track of causes; otherwise by going backwards we could reach an undefined state (where the extruder of a name is reversed, but not the action using that name in the subject position).

We consider two possibilities for keeping track of causes: the first one is by choosing one of the possible extruders and the second one is recording all of them. In the first case, we would obtain a notion of causality similar to the one introduced in~\cite{CristescuKV13}. In the following we shall concentrate on the second option. The idea is that, since we do not know which extruder really caused the action on an extruded name, we shall record the whole set of extruders that happened previously. In the framework, the set of executed extruders is set $\om$. The extrusions which are part of synchronisations will be deleted from $\om$ with the operation $\#$.

The predicates from the rules of
Figure~\ref{fig:ruleparam} are defined as follows: 

\begin{definition} [Disjunctive causality] If an indexed set $\g_{\om}$ is chosen as a data structure for a memory $\de$, the predicates are defined as:
\begin{description}
\item [1.] $\caus{\de,K,K'}=\caus{\g_{\om},K,K'}$ stands for $K'=K\cup \om$
\item [2.] $\upda{\de,K,K'}=\upda{\g_{\om},K,K'}$ stands for $K'=K$
\end{description}
\end{definition}
In the following example we shall give the intuition of how our framework captures the defined notion of causality.
\begin{example}\label{ex:yoshida}
Let us consider the process $X=\res a_{\emptyset_{\{*\}}} (\outin{b}{a^*}{*}\para \outin{c}{a^*}{*}\para a^{*}(x))$. By applying a rule $\textsc{Open}$ twice and executing concurrently two extrusions on names $b$ and $c$, we obtain a process:
$$\res a_{\{i,h\}_{\{*,i,h\}}} (\outin{b}{a^*}{*}\mem{i,*}\para \outin{c}{a^*}{*}\mem{h,*}\para a^{*}(x))$$
By definition of the predicate $\caus{\cdot}$,
the third action will take the whole index set $\{*,i,h\}$ as a set $K$ and we get the process:
$$\res a_{\{i,h\}_{\{*,i,h\}}} (\outin{b}{a}{*}\mem{i,*}\para \outin{c}{a}{*}\mem{h,*}\para a^{*}(x)\mem{l,\{*,i,h\}})$$
In the memory $\mem{l,\{*,i,h\}}$ we see that the first action to be reversed is the action with key $l$; the other two actions can be reversed in any order. 
\end{example}

\section{Properties}\label{sc:results}
In this section we shall show some properties of our framework. First we shall show that the framework is a conservative extension of standard $\pi$-calculus and that it enjoys causal consistency, a fundamental property for reversible calculi. After that, we shall prove causal correspondence between the causality induced by Boreale-Sangiorgi semantics and causality in the framework when $\de=\g_w$.
\begin{definition}[Initial and Reachable process]
A reversible process X is \emph{initial} if it is derived from a $\pi$-calculus process $P$ where all the restricting operators are initialised and in every prefix, names are decorated with a $*$. A reversible process is \emph{reachable}
if it can be derived from 
an initial
 process by using the rules in Figures~\ref{fig:rulecommon},~\ref{fig:ruleparam} and~\ref{fig:ruleback}.
\end{definition}

\subsection{Correspondence with the $\pi$-calculus}\label{ssc:correspondence}

We now show that our framework is a conservative extension of the $\pi$-calculus. To do so, we first define an erasing function $\f$ that given a reversible process $X$, by deleting all the past information, generates a $\pi$ process. Then we shall show that there is a \textit{forward} operational correspondence  between a reversible process $X$ and $\f(X)$. Let $\mathcal{P}$ be the set of $\pi$-calculus processes; then we have:

\begin{definition}[Erasing function]
The function $\f: \mathcal{\revprocs}\rightarrow\mathcal{P}$ that maps reversible processes to the $\pi$-calculus, is inductively defined as follows:
 \vspace{-1mm}
\begin{align*}
&\f(X\para Y)=\f(X)\para \f(Y)\quad &&\f(\hist [X])=\f(X)&&\f(\nil)=\nil\\
&\f(\res a_{\de} (X))= \f(X)\quad \quad\;\text{if } \empt(\de)=false \quad &&\f(b\ins{j}(x).\pj)=b(x).\f(\pj)\\
&\f(\res a_{\de} (X))= \res a \;\f(X)\quad \text{if } \empt(\de)=true\quad &&\f(\outi{b}{a^{j'}}.\pj)=\out{b}{a}.\f(\pj)
\end{align*}
The erasing function can be extended to labels as:
\vspace{-1mm}
\begin{align*}
&\f(\trans{i}{K}{j}{\pi})=\f(\pi)\quad &&\f(\out{b}{a})=\out{b}{a}\\
&\f(\outbo{b}{a}{\de})=\outbo{b}{a}{}\quad\text{when }\empt(\de)=true\quad &&\f(b(x))=b(x)\\
&\f(\outbo{b}{a}{\de})=\out{b}{a}\qquad\;\text{when }\empt(\de)=false\quad &&\f(\tau)=\tau
\end{align*}
\end{definition}
As expected the erasing function discards the past prefixes and name restriction operators that are non-empty. Moreover, it deletes all the information about the instantiators.

Every forward move of a reversible process $X$ can be matched in the $\pi$-calculus. To this end we use $\flts{}_\pi$ to indicate the transition semantics of the $\pi$-calculus.

\begin{lemma}\label{lm:fwsound}
If there is a transition $X\flts{\mu}Y$ then $\f(X)\flts{\f(\mu)}_{\pi}\f(Y)$.
\end{lemma}

We can state the converse of Lemma~\ref{lm:fwsound} as follows:
\begin{lemma}\label{lm:fwcomp}
If there is a transition $P\flts{\f(\mu)}_{\pi}Q$ then for all reachable $X$ such that $\f(X)=P$, there is a transition $X\flts{\mu}Y$ with $\f(Y)=Q$.
\end{lemma}

\begin{corollary}
The relation given by $(X,\f(X))$, for all reachable processes $X$, is a strong bisimulation.
\end{corollary}

\subsection{The main properties of the framework}\label{ssc:mainprop}
We now prove some properties of our framework which are typical of a reversible process calculus~\cite{rccs,CCSK,rhojournal,CristescuKV13}. Most of the terminology and the proof schemas are adapted from~\cite{rccs,CristescuKV13} with more complex arguments due to the generality of our framework.
The first important property is the so-called Loop Lemma, stating that any reduction step can be undone. Formally:

\begin{lemma}[Loop Lemma] \label{lem:loop}
For every reachable process $X$ and forward transition $t:X\flts{\mu}Y$ there exists a backward transition $t':Y\rlts{\;\mu}X$, and conversely. 
\end{lemma}

Before stating our main theorems we need to define the causality relation. It is defined on the general framework
 and it is interpreted as the union of \textit{structural} and \textit{object} causality. 
\begin{definition}[Structural cause on the keys]
For every two keys $i_1$ and $i_2$ such that $i_1 , i_2\in\key(X)$, we let $i_1 \stru_{X} i_2 $ if $X=C[\alpha[i_1,K_1].Y]$ 
and $i_2\in\key(Y)$.
\end{definition}
Once having defined structural causal relation on keys, we can extend it to transitions. 
%
\begin{definition}[Structural causality]
Transition $t_1 : X\lts{\trans{i_1}{K_1}{j_1}{\pi_1}} X'$ is a \emph{structural cause} of transition $t_2 : X''\lts{\trans{i_2}{K_2}{j_2}{\pi_2}} X'''$, written $t_1\stru t_2$, if 
$i_1 \stru_{X'''} i_2 $ or $i_2 \stru_{X} i_1 $. \emph{Structural causality}, denoted with $\str$, is the reflexive and transitive closure of $\stru$.
\end{definition}

Object causality is defined directly on the transitions and to keep track of it we use the contextual cause $K$.
\begin{definition} [Reverse transition]
The reverse transition of a transition $t: X\flts{\mu} Y$, written $t\rev$, is the transition
%
 with the same label and the opposite direction $t\rev: Y\rlts{\mu}X$, and vice versa. Thus $(t\rev)\rev = t$.
\end{definition}
\begin{definition}[Object causality]\label{def:object}
Transition $t_1 : X\lts{\trans{i_1}{K_1}{j_1}{\pi_1}} X'$ is an \emph{object cause} of transition $t_2 : X'\lts{\trans{i_2}{K_2}{j_2}{\pi_2}} X''$, written $t_1\ob t_2$, if 
$i_1\in K_2 $
or 
$i_2\in K_1 $
(for the backward transition) and $t_1\ne t_2\rev$.
\emph{Object causality}, denoted with $\obj$, is the reflexive and transitive closure of $\ob$.
\end{definition}
\begin{example} 
Consider a process $X=\res a_{\de} (\outin{b}{a^*}{*}\para \outin{c}{a^*}{*}\para a^*(z))$ and the case when $\de=\emptyset_*$, as in Example~\ref{ex:boreale}. The executed actions would be $\lts{\trans{i_1}{*}{*}{\outbo{b}{a}{\emptyset_*}}}\lts{\trans{i_2}{\{i_1\}}{*}{\outbo{c}{a}{{\{i_1\}}_{i_1}}}}\lts{\trans{i_3}{\{i_1\}}{*}{a(z)}}$. We can notice that $K_2=\{i_1\}$ and $K_3=\{i_1\}$, indicating that the second and the third action are caused by the first one. By choosing a different data structure we can obtain different causal order, as mentioned in Example~\ref{ex:ex1} and Example~\ref{ex:yoshida}.
\end{example}
\vspace{1mm}
\begin{definition}[Causality relation and concurrency]
The causality relation $\cau$ is the reflexive and transitive closure of structural and object cause:
 $\cau = (\str\cup\obj)^{*}$. Two transitions are \emph{concurrent} if they are not causally related.
\end{definition}

Concurrent transitions can be permuted, and the commutation of transitions
is preserved up to label equivalence. 
\vspace{1mm}
\begin{definition}[Label equivalence]
Label equivalence, $=_{\lambda}$, is the least equivalence relation satisfying: $\trans{i}{K}{j}{\outbo{b}{a}{\de}}=_{\lambda}\trans{i}{K}{j}{\outbo{b}{a}{\de'}}$ for all $i, j, K, a ,b$ and $\de, \de' \subseteq \ids$.
(Having an indexed set $\g_w$ for $\de$ we disregard index $w$, and observe 
$\g \subseteq \ids$.)
\end{definition}
\vspace{1mm}
\begin{lemma}[Square Lemma]\label{lm:square}
If $t_1 : X\lts{\mu_1} Y$ and $t_2 : Y\lts{\mu_2} Z$ are two concurrent transitions, there exist  $t'_2 : X\lts{\mu'_2} Y_1$ and $t'_1 : Y_1\lts{\mu'_1} Z$ where $\mu_i=_{\lambda}\mu'_i$. 
\end{lemma}
We shall follow the standard notation and say that $t_2$ is a residual of $t'_2$ after $t_1$, denoted with $t_2=t'_2/t_1$.
Two transitions are \textit{coinitial} if they have the same source, \textit{cofinal} if they have the same target, and \textit{composable} if the target of one is the source of the other. A sequence of pairwise composable transitions is called a \textit{trace}, written as $t_1; t_2$. We denote with $\epsilon$ the \textit{empty} trace.
Notions of target, source, composability and reverse extend naturally to traces.

With the next theorem we prove that reversibility in our framework is causally consistent.
 \begin{definition}[Equivalence up-to permutation]\label{df:equiv}
Equivalence up-to permutation, $\sim$, is the least equivalence relation on the traces, satisfying:
\vspace{-3mm}
$$t_1;(t_2/t_1)\eq t_2;(t_1/t_2)\qquad t;t\rev\eq\epsilon$$
\end{definition}
\vspace{-1mm}
Equivalence up-to permutation introduced in~\cite{rccs} is an adaptation of equivalence between
traces introduced in~\cite{levy,castellani} that additionally erases from a trace, transitions triggered in both directions. It just states that concurrent actions can be swapped and that a trace made by a transition followed by its inverse is equivalent to the empty trace.

\begin{theorem}\label{th:equivalent}
Two traces are coinitial and cofinal if and only if they are equivalent up-to permutation.
\end{theorem}

\subsection{Correspondence with Boreale and Sangiorgi's semantics}\label{ssc:bscorrespondence}

We prove causal correspondence between Boreale and Sangiorgi's late semantics (rather than early, as originally given) and the framework when memory $\de$ is instantiated with $\g_w$. The precise definitions and the proofs are given in~\cite{MMPY18};
here we shall give just a brief presentation of the idea.

To compare semantics, we observe traces (runs) of the processes.
Labels in the framework will bring additional information about the multiset of the structural causes ($K_{F}$) of the executed action and a trace in the framework will have the following form: $X_1\flts[K_{F1}]{\mu_1}X_2\ldots X_{n-1}\flts[K_{Fn}]{\mu_n}X_{n}$. 
To simplify notation, we shall write the transition $A\ltss{K;k}{\pi}A_2$ from Boreale and Sangiorgi's semantics as $A\ltss{K_{B}}{\zeta}A_2$, where $\zeta=k: \pi$.

Focusing on the structural correspondence, the main difference is in the silent actions. In the framework,
silent actions are identified with unique keys, while in Boreale and Sangiorgi's semantics, they just merge the cause sets of the actions participating in the communication.
Hence, we need to provide a connection between sets of structural causes in these two semantics. Let us briefly explain our method; more details can be found in~\cite{MMPY18}.

Suppose that we have two transition $t$ and $t'$, where $t: X\flts[K_{F}]{\trans{i}{K}{j}{\pi}}X'$ and $t': A\ltss{K_B}{i:\pi}A'$ where the continuation of the processes $X$ and $A$ is $\pi.P$\footnote{By abuse the notation, we shall write $\pi$ for the prefixes and the labels of the actions in both semantics, since they are essentially the same}. 
We can represent the dependences between the keys in the history of the process $X$ (all executed actions in $X$) with a directed graph, in the following way: keys of executed actions will be represented as vertices of a graph (actions which are part of a communication and have the same key, will be represented by two vertices with the same name); order between keys will be represented with directed edges where between the same vertices we shall have edges in both directions. Let as denote this graph $G=(V,E)$, where $V$ is a multiset of vertices and $E$ set of edges. 
%

To show exact correspondence between cause sets $K_{F}$ and $K_B$ we need to take the history part of the process $X$ involved in the execution of an action $\pi$. We can do it by taking all the paths in $G$ in which the target vertex will be key $i$ of the action $\pi$ and we shall obtain the graph $G(i)=(V(i),E(i))$. The multiset $V(i)$ contains all the keys which cause the action $\pi $ including $i$ and we can conclude that $K_F=V(i)\setminus \{i\}$. 
By removing all bidirectional edges from the graph $G(i)$ and replacing vertices that they connect with vertex renamed to $\tau_l$ when $l=1,2,\ldots n$, we shall obtain the graph $G'=(V',E')$. (Renaming is applied also on the other edges containing removed vertices. The operation of bidirectional edge contraction is precisely defined in~\cite{MMPY18}.)
The set $V'$ differs from $V(i)$ in having $\tau_l$ vertices instead of the pairs of the vertices with the same name (originally belonging to silent moves in the framework). Hence, we can conclude that
 $K_B=V'\setminus (\{i\}\cup \tau_l)$.

We shall call the algorithm explained above `Removing Keys from a Set', denoted as $\mathtt{Rem}$. We shall write $\mathtt{Rem}(K_{F})=K_B$, meaning that $K_B$ can be obtained by applying algorithm $\mathtt{Rem}$ to $K_F$.


Before stating the theorem, we shall give a definition of the erasing function $\lambda$ and the function $\gamma$ that maps labels from the framework into labels from Boreale and Sangiorgi's semantics:
\begin{definition}\label{def:mapping}
 The function $\gamma$ that maps label from the framework with a label from Boreale and Sangiorgi's semantics, is inductively defined as follows:
 \vspace{-2mm}
\begin{align*}
&\gamma(\trans{i}{K}{j}{\pi})=i:\gamma(\pi)\quad \quad\text{when }\pi\neq\tau\quad&&\gamma(\trans{i}{*}{*}{\tau})=\tau\\
&\gamma(\outbo{b}{a}{\de})=\outbo{b}{a}{}\quad\text{when }\empt(\de)=true\quad &&\gamma(b(c))=b(c)\\
&\gamma(\outbo{b}{a}{\de})=\out{b}{a}\qquad\;\text{when }\empt(\de)=false\quad&&\gamma(\out{b}{a})=\out{b}{a}
\end{align*}
\end{definition}
\begin{definition} \label{def:erasingbs}
The erasing function $\lambda$ that maps causal processes from Boreale and Sangiorgi's semantics to the $\pi$-calculus is inductively defined as follows:
 \vspace{-2mm}
\begin{align*}
&\lambda(A\para A')=\lambda(A)\para \lambda(A')&&\lambda(\mathtt{K}::A)=\lambda(A)&
&\lambda(\res a (A))=\res a (\lambda(A))&&\lambda(P)=P&
\end{align*}
\end{definition}

Now we have all necessary definitions to state the lemma about structural correspondence between two causal semantics. 

\begin{lemma}[Structural correspondence] \label{the:strcorrespondence}
Starting from initial $\pi$-calculus process $P$, where $P=A_1=X_1$, we have:
 \vspace{-2mm}
 \begin{enumerate}
\item if $P\ltss{K_{B1}}{\zeta_1}A_2\ldots A_{n}\ltss{K_{Bn}}{\zeta_{n}}A_{n+1}$ then there exists a trace $P\flts[K_{F1}]{\mu_1}X_2\ldots X_{n}\flts[K_{Fn}]{\mu_{n}}X_{n+1}$ and $K_{Fi}$, such that for all $i$, $\lambda(A_{i})=\f(X_{i})$, $\zeta_i=\gamma(\mu_i)$ and $\mathtt{Rem}(K_{Fi})=K_{Bi}$, for $i=1,...,n$.
\item if $P\flts[K_{F1}]{\mu_1}X_2\ldots X_{n}\flts[K_{Fn}]{\mu_n}X_{n+1}$ then there exists a trace $P\ltss{K_{B1}}{\zeta_1}A_2\ldots A_{n}\ltss{K_{Bn}}{\zeta_{n}}A_{n+1}$ where for all $i$, $\lambda(A_{i})=\f(X_{i})$, $\zeta_i=\gamma(\mu_i)$ and $\mathtt{Rem}(K_{Fi})=K_{Bi}$, for $i=1,...,n$.
\end{enumerate}
\end{lemma}
Considering the object dependence we have that the first action which extrudes a bound name will cause all 
future actions using that name in any position of the label. 
The main difference
 is that object dependence induced by input action in Boreale and Sangiorgi's semantics is subject dependence as well.

The next theorem demonstrates causal correspondence between causality in the framework when memory $\de$ is instantiated with $\g_w$ and Boreale and Sangiorgi's late causal semantics.
\begin{theorem}[Causal correspondence] \label{the:causalcorresp}
The reflexive and transitive closure of the causality introduced in~\cite{BorealeS98} coincides with the causality of  the framework when $\de=\g_w$.
\end{theorem}

\section{Conclusions} \label{sc:conc}
In a concurrent setting, causally-consistent reversibility relates causality and reversibility.
Several works~\cite{CristescuKV13,DeganoP99,BorealeS98,CrafaVY12,BusiG95} have addressed causal semantics for $\pi$-calculus, differing on how object causality is modelled.
Starting from this observation, we have devised a framework for reversible $\pi$-calculus which abstracts away from the underlying data structure used to record causes and consequences of an extrusion, and hence from 
the object causality.
Depending on the underlying data structure, we can obtain different causal semantics. We have shown how three different semantics~\cite{CristescuKV13,BorealeS98,CrafaVY12} can be derived, and we have proved causal correspondence with the semantics introduced in~\cite{BorealeS98}.
Our framework enjoys typical properties for reversible process algebra,
such as loop lemma and causal consistence.
 As a future work we plan to prove causal correspondence with the semantics~\cite{CristescuKV13,CrafaVY12}
and to continue working towards a more parametric framework and to
  compare it with~\cite{pereracheney2017,HildebrandtJN17}. Moreover it would be interesting to implement our framework in the psi-calculi framework~\cite{abs-1101-3262}, and to develop further 
  the behavioural theory of our framework.

 {\small
\section*{Acknowledgments}
We are grateful to the EXPRESS/SOS reviewers for their useful remarks and suggestions which led to substantial improvements.
}

\bibliographystyle{eptcs}
 \bibliography{biblio}
 





\end{document}